# Super Resolution Convolutional Neural Network for Feature Extraction in Spectroscopic Data

Han Peng[1], Xiang Gao[2], Yu He[3], Yiwei Li[1], Yuchen Ji[4], Chuhang Liu[4], Sandy A. Ekahana[1], Ding Pei[1], Zhongkai Liu[4], Zhixun Shen[3], Yulin Chen[1,4*]

[1]Clarendon Laboratory, Department of Physics, University of Oxford, Oxford, Oxfordshire, OX1 3PU, United Kingdom

[2]Department of Chemistry, University of Florida, Gainesville, Florida, FL 32603, United States

[3]Department of Applied Physics, Stanford University, Stanford, California, CA 94305, United States

[4]School of Physical Science and Technology, ShanghaiTech University and Chinese Academy of Sciences-Shanghai Science Research Center, Shanghai, 201210, China

*Corresponding author: Prof Yulin Chen, yulin.chen@physics.ox.ac.uk

## Abstract

**Two dimensional (2D) peak finding is a common practice in data analysis for physics experiments, which is typically achieved by computing the local derivatives. However, this method is inherently unstable when the local landscape is complicated, or the signal-to-noise ratio of the data is low. In this work, we propose a new method in which the peak tracking task is formalized as an inverse problem, thus can be solved with a convolutional neural network (CNN). In addition, we show that the underlying physics principle of the experiments can be used to generate the training data. By generalizing the trained neural network on real experimental data, we show that the CNN method can achieve comparable or better results than traditional derivative based methods. This approach can be further generalized in different physics experiments when the physical process is known.**

## Introduction

Recent advances on experimental techniques in condensed matter physics boost the generation of large volume high-quality data. In order to present these data, 2D (and even 3D) representations of the data becomes more and more popular, such as in angular resolved photo-emission spectroscopy (ARPES)[1], scanning tunneling microscopy (STM)[2], resonant inelastic x-ray scattering (RIXS)[3], etc. In these experiments, as the data quality is often limited by the instrumental resolution and different intrinsic physical processes, the retrieving of physical quantities with high precision can therefore benefit from effective data analysis methods.

As an example, a typical 2D ARPES experiment data set is shown in Fig. 1. Ideally, the ARPES spectra should follow the theoretical energy-momentum dispersion shown in Fig. 1A. However, intrinsic broadening effects[1] such as electronic correlation, as well as extrinsic factors such as crystal defects can broaden the spectrum in both energy and momentum dimensions. Together with the resolution limitation, sometimes it is difficult to resolve the energy bands in the 2D measurements (e.g. in Fig. 1B). To enhance the features in the 2D band dispersions, several derivative based methods have been proposed, such as the Maximum Curvature (MC) method[4] and the Minimum Gradient (MG) method[5]. The MC method calculates the local curvature and assumes the pixels with large curvature are the positions of the energy bands; the MG method calculates the average gradient and assumes the positions with small average gradient represent the energy band location. However, these methods can only give reasonable results in high signal-to-noise ratio data and tend to fail in complicated situations when multiple bands are close to each other or when the data are too noisy.

On the other hand, recent development of machine learning and deep learning techniques such as convolutional neural network (CNN) provide great performance in improving 2D data qualities by solving a series of inverse problems, such as super-resolution, denoising and patching[6-8]. As high-quality 2D data (i.e. images) are subjected to various degrading transformations in experiments, the objective of our data analysis becomes finding an appropriate inverse transformation that can best recovers the original images – which can be formulated as solving an inverse problem.

Motivated by this insight, we consider the energy band extraction problem in a 2D ARPES image (broadened by the intrinsic and extrinsic processes as discussed above) as a problem of inversing the physical processes that blurs the spectral peaks along the band dispersions, and thus it can be treated using CNN. In other words, we effectively look for a map between the "broadened experimental dispersions" and the "original dispersions". Moreover, leveraging the existing knowledge of the intrinsic and extrinsic broadening processes, massive simulated data can be generated for the training purpose before we apply the trained model to the real experimental data.

In this work, we use the simulated ARPES data to train a modified Super-Resolution Convolutional Neural Network (SR-CNN) that fits the ARPES experiments' spectral intensity map with the extrema feature map, thus visualizes the positions of the energy bands. SR-CNN was originally proposed for mapping low resolution image patches to corresponding high resolution patches, which is an inverse problem of down sampling in image processing.[9] Comparing to other recent neural network architectures in solving this problem that uses very deep net[7, 8], the SR-CNN has only three layers and the function of each layer can be well understood. We demonstrate that this method can resolve complicated features in ARPES data and outperforms previous traditional algorithms in noise-resilience, sharpness and accuracy.

## Method and Results

### Results from different algorithms

For reference, the theoretically calculated energy band is presented at Fig. 1A to compare with the experiment data in Fig. 1B. The key feature showing in the calculation is that there are three energy bands crossing the Fermi energy ($E_F$) and the band No.2 terminates around $E_F$ and merges with other faint bands. In addition, band No. 1 has a degenerate point at k = 0 about 60meV below the $E_F$. Ideally, a successful data analysis algorithm should resolve all these features from the ARPES intensity map.

Due to the experiment restriction, the number of pixels in the current data along k-direction is only 29. The low resolution makes it even more difficult to directly track the features from the experiment data. The MC and MG methods show significant enhancement of the energy band features (as shown in Fig. 1 C and D) but are either noisy (MC) or still blurry (MG). The result of our method is shown in Fig. 1E, demonstrating all three $E_F$-crossing bands and the crossing below FS. It also shows that the band in the middle (No.2) ends at about -15meV below the FS, which matches the theoretical calculation.

### Super-resolution convolutional neural network

The design of our CNN method is modified from SR-CNN, a three-layer fully convolutional neural network proposed by Dong, Loy, He and Tang for image super-resolution.[9]

The architecture of our modified SR-CNN is shown in Fig. 2A. In SR-CNN, each layer is understood as an operation that is comparable to traditional image restoration methods[6]: The first layer transforms the raw experimental image patches to an $n_1$-dimensional representation under a trained basis set. The second layer non-linearly maps the new representation to the $n_2$-dimensional sharp energy band feature space. At the final stage, the image of energy band is reconstructed by averaging all the features from the adjacent area.

Assuming **Y** is the input image, $W_i$ and $B_i$ are the weights and bias in the $i$-th layer, this three-layer neural network can be mathematically written as:

$$F_1(\mathbf{Y}) = \max(0, W_1 * \mathbf{Y} + B_1),$$

$$F_2(\mathbf{Y}) = \max(0, W_2 * F_1(\mathbf{Y}) + B_2),$$

$$F(\mathbf{Y}) = \text{sigmoid}(W_3 * F_2(\mathbf{Y}) + B_3)$$

The '*' denotes for a convolution. The result of the convolution in the first two layers is passed to a rectified linear unit ($\text{ReLU}(x) = \max(0, x)$) to produce non-linearity[10]. The output layer is with a sigmoid activation function ($\text{Sigmoid}(x) = \frac{1}{1+e^{-x}}$).

Note that although the architecture we are using is almost the same as SR-CNN, the problem we are trying to solve is different from what SR-CNN was originally designed for: In the original SR-CNN, the super-resolution problem was formulated as a regression problem, where no non-linear activation function was applied in the last layer, and the mean squared error (MSE) loss between the network output and the ground truth was minimized. In our design, the problem was formulated as $h_{out} \times w_{out}$ classification problems. The sigmoid activation function is applied in the third layer to obtain the confidence for the existence of the energy band, giving an output range from 0 to 1 naturally. We minimize the standard loss function for classification, i.e. the cross entropy, also known as the negative log likelihood of the ground truth, which is defined as

$$\mathcal{L}(\mathbf{X}, \widehat{\mathbf{X}}; \boldsymbol{\theta}) = -\frac{1}{N} \sum_{n=1}^{N} [\mathbf{X}_n \log \widehat{\mathbf{X}}_n + (1 - \mathbf{X}_n) \log(1 - \widehat{\mathbf{X}}_n)],$$

where $\mathbf{X}_n$ denotes for each pixel in the label, $\widehat{\mathbf{X}}_n$ for SR-CNN output from the training data, and $\boldsymbol{\theta}$ denotes for the parameters in the SR-CNN (i.e. bias and weights).

The setting we used is $n_1 = 64, f_1 = 9$ for $\text{size}(W_1) = [n_1, f_1, f_1, 1]$ ; $n_2 = 32$ for $\text{size}(W_2) = [n_2, 1, 1, n_1]$; $f_3 = 5$ for $\text{size}(W_3) = [1, f_3, f_3, n_2]$, which is default as shown in the previous SR-CNN work[6]. The output size can be derived as:

$$h_{out} = h_{in} - (f_1 - 1) - (f_2 - 1) - (f_3 - 1) = h_{in} - 12$$

$$w_{out} = w_{in} - (f_1 - 1) - (f_2 - 1) - (f_3 - 1) = w_{in} - 12$$

The receptive field of each output pixel is 13-by-13 (i.e. each output pixel is determined locally by the neighboring $13 \times 13$ pixels).

### The training processes

For a low-level vision task such as super-resolution, the training process follows the paradigm of self-supervised learning which does not require to manually label the training data. The low-resolution inputs **Y** are generated by down-sampling the high-resolution image patches **X** in the training set. The output $\widehat{\mathbf{X}}$ generated from the forward propagation process is then compared with the original high-resolution image **X** to compute the loss function $\mathcal{L}(\mathbf{X}, \widehat{\mathbf{X}}; \boldsymbol{\theta})$. Backpropagation is used to minimize the loss function $\mathcal{L}$ with respect to $\boldsymbol{\theta}$ iteratively[9]. In this process, the high-resolution image patches are used as the label, and the down-sampled ones are used as the input of the model.

For the task of the energy band feature extraction from the ARPES experiment data, the training data and the label should be chosen more carefully. For any experiment data, it is not possible to label the energy band position automatically and it is not accurate to label it by human guess. To solve this dilemma, simulated data are used for the training and testing process while the experiment data are used to evaluate the result.

Fig. 2B shows the data simulation and the training process. The energy band $\epsilon(\mathbf{k})$ is generated through a tight binding model with randomized parameters. Since the SR-CNN output is only affected locally by the neighboring 13-by-13 pixels, the global property such as symmetry of the energy band does not affect the model. The energy band image is then generated by discretizing $\epsilon(\mathbf{k})$ through putting 1's in the occupied pixels and 0's in the unoccupied pixels. The generated image is finalized by super-sampling for anti-aliasing[11] and then used as the label.

To simulate the detected experimental data corresponding to the label, the ARPES spectrum broadening model is used. Assuming we have an energy band with a shift due to self-energy: $E = \epsilon(\mathbf{k}) + \Sigma'(\mathbf{k}, E)$, the one-particle spectral function is written as[1]

$$A(\mathbf{k}, E) = -\frac{1}{\pi} \frac{\Sigma''(\mathbf{k}, E)}{\left(E - \epsilon(\mathbf{k}) - \Sigma'(\mathbf{k}, E)\right)^2 + \Sigma''(\mathbf{k}, E)^2}$$

which means the energy band is broadened and shifted by the self-energy that comes from the electronic interactions. And the observed data, the photon count, follows

$$I_0(k, E) \sim \left\{ A(k, E) FD(E) |M(k)|^2 + g_n^{in}(k, E) \right\} \otimes R(\Delta k, \Delta E) + g_n^{ex}(k, E),$$

where $FD(E)$ is Fermi-Dirac distribution, $M(k)$ is the matrix element, $g_n^{in}(k, E)$ is the intrinsic noise such as the shot noise, $\otimes$ denotes convolution, $R(\Delta k, \Delta E)$ is the instrument resolution and $g_n^{ex}(k, E)$ is the extrinsic noise such as circuit noise.

When the variation of self-energy cannot be ignored within the convolution window in the targeting experiment data, then this effect should be considered in the simulation process. However, as a proof-of-concept design, we currently assume this part is slow-variate and can be ignored in our model due to the localized, all-convolution network design. To generate the simulated data, $\Sigma'$ is not considered since as the result is interpreted as the local peak positions in the experimental data. The energy band image is convolved with the one-particle spectral of randomized self-energy $\Sigma''$ and is added with a Poisson noise. Gaussian blurring is then applied to simulate the instrument resolution. A Gaussian noise is added at the final step for the circuit noise. The finalized broadened and noisy simulated image is used as the training data.

The optimization is done iteratively. Before the training process, multiple (~20 in the presented study) simulated 3D band structures are generated, and 1000 2D-slices are cut from the volume as training data. The initial value of the filters and bias are from the result of the SR-CNN in natural images[6]. During one training step, multiple patches randomly selected from the training data are passed through the forward propagation to produce the output. The loss function $\mathcal{L}$ is then calculated between the output and the label to produce a gradient flow to determine the changing direction of parameters. An Adam optimizer is used for the backpropagation[12] to update the parameters. The training is iterated for ~30 million steps until the parameters the model performance saturates.

# Discussion

## Benchmark of the performance

To further compare the performance and the reliability of different data analysis methods, we apply each method to a variety of data sets, with the result presented in Fig. 3 and Table 1.

Fig. 3 A and B compare the sensitivity to band gaps for different methods. We applied each method to a simulated data set with a small energy gap of ~30meV. The gap is indistinguishable either from the visual inspection of the image or from the energy distribution curve (EDC). The results of different methods are summarized in Table 1. The MC method increases the visibility of the gap feature with an inferred gap size of 25meV from fitting, but the output is noisy (Fig. 3 A.iii and B.iii). The Minimum Gradient resolves most of the band clearly but falsely produces high intensity values inside the gap (Fig. 3 A.iv and B.iv). This is because the local intensity landscape is a saddle point (minima long E direction but maxima along k direction), where the average gradient reaches minimum as it does at a peak point or at the ridge. This saddle-point problem prevents MG method from resolving small gap feature in general. As shown in Fig. 3 A.v and B.v, CNN resolves the gap with a size of 24meV. Though MC and CNN provide the comparable results in gap size, the contrast of the CNN result is much larger, and the resolved bands are sharper. Also, the result is more noise resilient in CNN.

To evaluate the performance of our algorithm in experiment data, an example of the 2D experimental dispersion (from sample PtSe$_2$ in Ref. 13) is used in Fig. 3C. A degeneracy point is expected in the PtSe$_2$ band structure (Fig 3C.i). The experiment data is too blurry to directly delineate the two bands (Fig. 3C.ii); and neither the MC method or the MG method can resolve the two bands clearly (see Fig. 3C. ii and iii). However, the CNN method can produce clear result showing the two bands crossing with less noise and artefacts (Fig. 3C.v). This result show that the CNN method can achieve comparable or better performance than the traditional methods, especially in the resilience to the noise and sharpness of the resulting features.

While CNN method extracts the main feature in Fig. 3C.v successfully, one can notice an unexpected up-bending feature at $k\sim0.06\text{Å}^{-1}$ near $E_D$, which is not shown in the computational result. To test whether this is a new feature or an artifact, we corrupt the data with gaussian noise in Fig. 3D. While the CNN results with the corrupted data show consistent features of the band crossing, there is no sign of the up-bending band. Thus, we can infer that the unexpected feature may be an artifact accidentally introduced by experimental noise or pre-processing steps.

The consistent results with noise-corrupted data show the robustness of the CNN method. However, the result quality reduces qualitatively as the noise goes up. Therefore, one should be cautious when interpreting the CNN results for experiments with high signal-to-noise ratio and repeat the experiment when necessary.

## Understanding the CNN model

To understand the mechanism of the CNN method, the filters and responses from the first layer are visualized in Fig. 4. Although not all filters have clearly understandable meanings, some of the filters can be identified with specific function. Some of the filters are sensitive to the inclined ridges (e.g. filters No. 1 and 2); some of the filters are sensitive to flat bands (e.g. filter No. 3); some of the filters act like 'background detector', which separate the band and the noisy background (e.g. filter No.4); and some of the filters act as a convolution along the horizontal axis (e.g. filter No. 5), which is similar to the traditional energy-momentum curve peak finding method[4]. In the second layer, these feature maps are averaged with weights and finally used to reconstruct the sharp-feature map in the output layer.

The output of the CNN method marks the position of the peaks, which stands for the position of the energy band. However, the information of the self-energy, which is of interest in analyzing the behavior of the correlated system[14, 15], is not extracted. In the future study, this part of information could be included by adding a branch in the neural network.

### Overfitting

One issue of using neural network is over-fitting[16], when the training set is too small and the model fails to generalize to the other data sets. This issue should be tackled carefully when using the simulated data, as shown in Fig. 5. Since the experiment data is always noisy, a model trained with the noise-free simulated data can treat the experimental noise as separated peaks, resulting in discontinuity of the output energy band, as shown in Fig. 5B (marked with the red arrow 1). Smoothing is commonly used as a pre-processing step to suppress the noise and enhance the continuity of the output energy bands in both MC and MG methods, and it can also be used before applying CNN, as shown in Fig. 5C. However, smoothing also causes loss of small features (as marked with the red arrow 2 in Fig. 5, B and C). To avoid the smoothing step and enhance the model's stability under noise, random artificial noise is added in the training data. As a result, the output of the model for the experiment data shows a continuous energy band without necessity of smoothing and thus keeps the detailed features (Fig 5D).

### Generalizability

The training data generation process only considers the one-particle spectral function and the Fermi-Dirac distribution. The self-energy is randomized and fixed as constant for any single simulated data, and is varied between 10 meV and 45 meV in the training set. Although this range only covers limited values for the self-energy, the experimental data can be resampled in the analysis so that the peak width matches the training set in the unit of pixel. In this case, we deliberately choose to have this proof-of-principle model free from any other physical effect explicitly. However, if the self-energy variation is slow enough (i.e. near constant within 13-by-13 pixels), the model is still expected to robustly work as an approximation, due to the independent nature among each patch fed into the CNN convolutional window. The detected peak position in such systems may be systematically drifted by what one would expect from EDC/MDC fitting.

We demonstrate this effect by evaluate our CNN model in the Bi2212 system (Fig. 6). The nodal spectra we use here is chosen from deeply underdoped side of the phase diagram (p = 0.063, Tc = 22K)[17, 18], where both electron-electron correlation and electron-phonon coupling effects are universally accepted as strong. In particular, the electron self-energy has both a highly binding energy-dependent component due to correlations, as well as a characteristic phonon-kink at around $E_B \sim 70$ meV. Traditional way of deciding the dispersion involves either EDC or MDC fitting, which tend to have artefacts when the dispersion is heavily interaction-broadened and too steep/flat. One can see that, the CNN successfully extracted the 'kink' in the band structure, which, even better, lies right in between the MDC (red line)/EDC (red cross) results. This shows the advantage of the 2D nature of the CNN method, which overcomes unwanted biases introduced in unidirectional analysis such as 1D EDC/MDC fitting or unidirectional second derivative method. For future studies, one could expand the training assumptions and incorporate these physical processes in training data generation to achieve more accurate results.

Our model shows good performance in the presented datasets. However, one would expect for more complex datasets (e.g. with large noise, complex local geometry or unmodelled physical properties), well-designed preprocessing steps will be needed, and the model may need further improvement.

To conclude, in this work, we demonstrate that the neural network method shows good sensitivity to the dispersive features in real ARPES data and its resilience to noise. Our results show that the feature extraction in physics experiment's 2D data analysis can be treated as an image restoration problem in computer vision and can be tackled with convolutional neural network. The training process which uses simulated data provides an example that the neural network can be used to solve the inverse problem when the physics model is known. Moreover, this methodology can also be used in general peak hunting from the multidimensional data and can be potentially used in other physical processes.

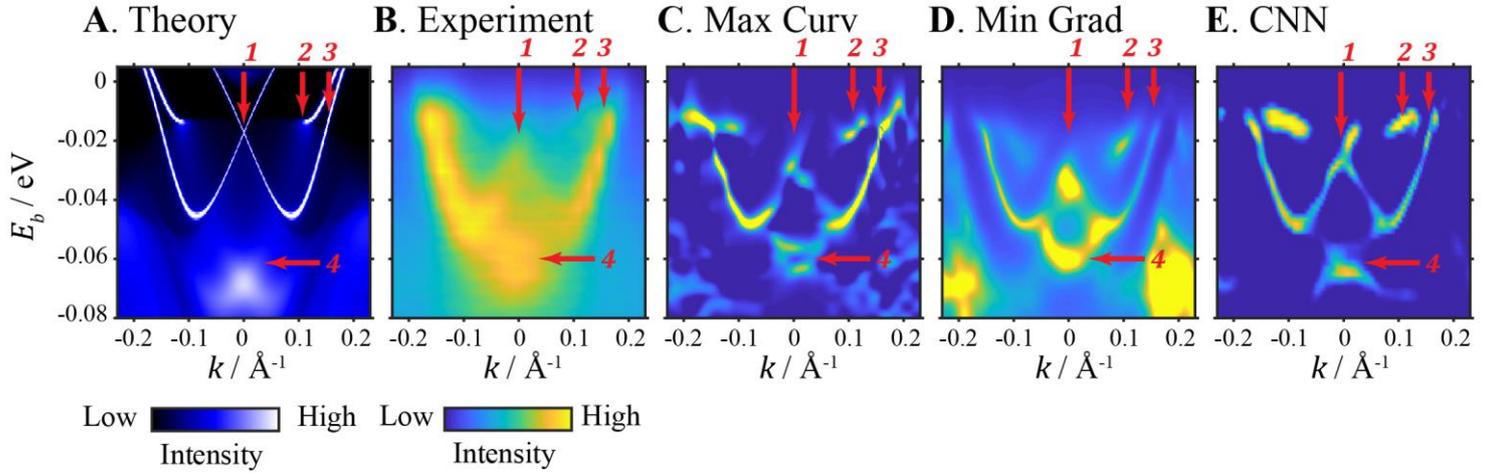

**Fig. 1. Comparison of experiment data, DFT calculation and different data analysis methods. (A)** Density functional theory calculation of the band structure of NbIrTe4. The three bands crossing the Fermi surface (Eb = 0) are marked with red arrows. The fourth arrow marks the crossing below the Fermi surface. The brighter color shows higher intensity. **(B)** The corresponding experiment data of the band structure. The data is used to produce the subsequent results in (C) – (E). The four red arrows mark the corresponding features in (A), which is blurry in the experiment. The yellow color shows high intensity while the blue color shows the low. **(C)-(E)** The result of Maximum Curvature (C), Minimum Gradient (D), and the CNN (E) methods that extract the energy band from the experiment data in (B). Interpolating and smoothing are used for pre-processing. As shown in the color bars, the colormap shows the intensity of the experiment data or the relative intensity of the algorithm output. This convention is used for the rest of the manuscript unless otherwise specified.

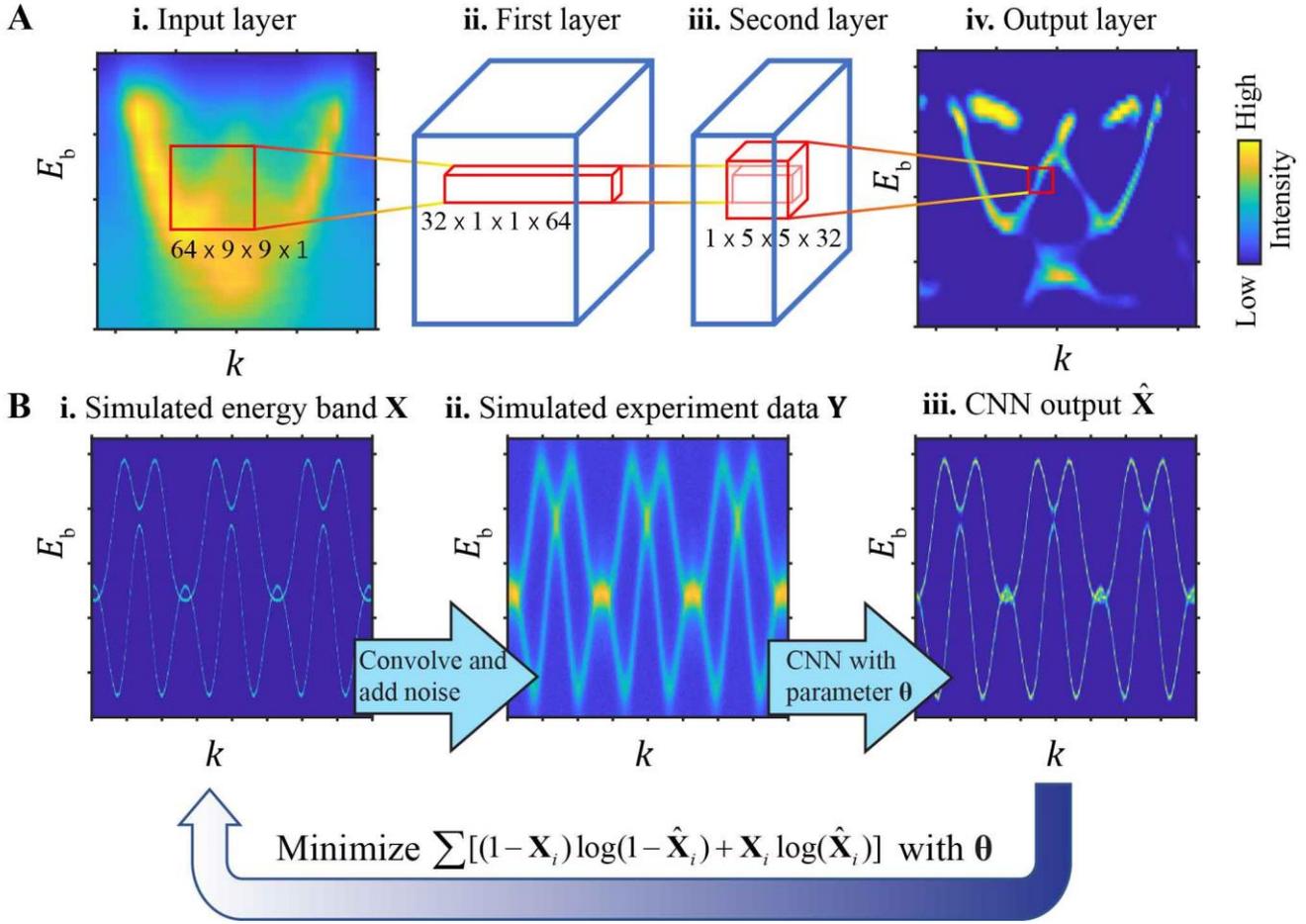

**Fig. 2. The architecture of the super-resolution convolutional neural network. (A)** The forward propagation from the experiment data (blurry) to the energy band feature map (sharp). The input layer: an experimental energy-momentum intensity distribution map. The first layer: convolution of the input layer with 64 filters with the size 9-by-9-by-1 with the ReLU activation function. The second layer: convolution with 32 filters with size 1-by-1-by-64 filters with the ReLU activation function. Output layer: convolution with one 5-by-5-by-32 filter with the Sigmoid activation function. **(B)** The training process using simulated data. **i)** The training label **X** is generated by a random-parameterized tight binding model. **ii)** The training data **Y** is produced by Lorentzian and Gaussian convolution of the label with random noise, and then used as the input of the CNN. **iii)** The CNN output $\hat{\mathbf{X}}$ is compared with the label **X** to calculate the gradient. The loss function is minimized in the iteration based on the gradient.

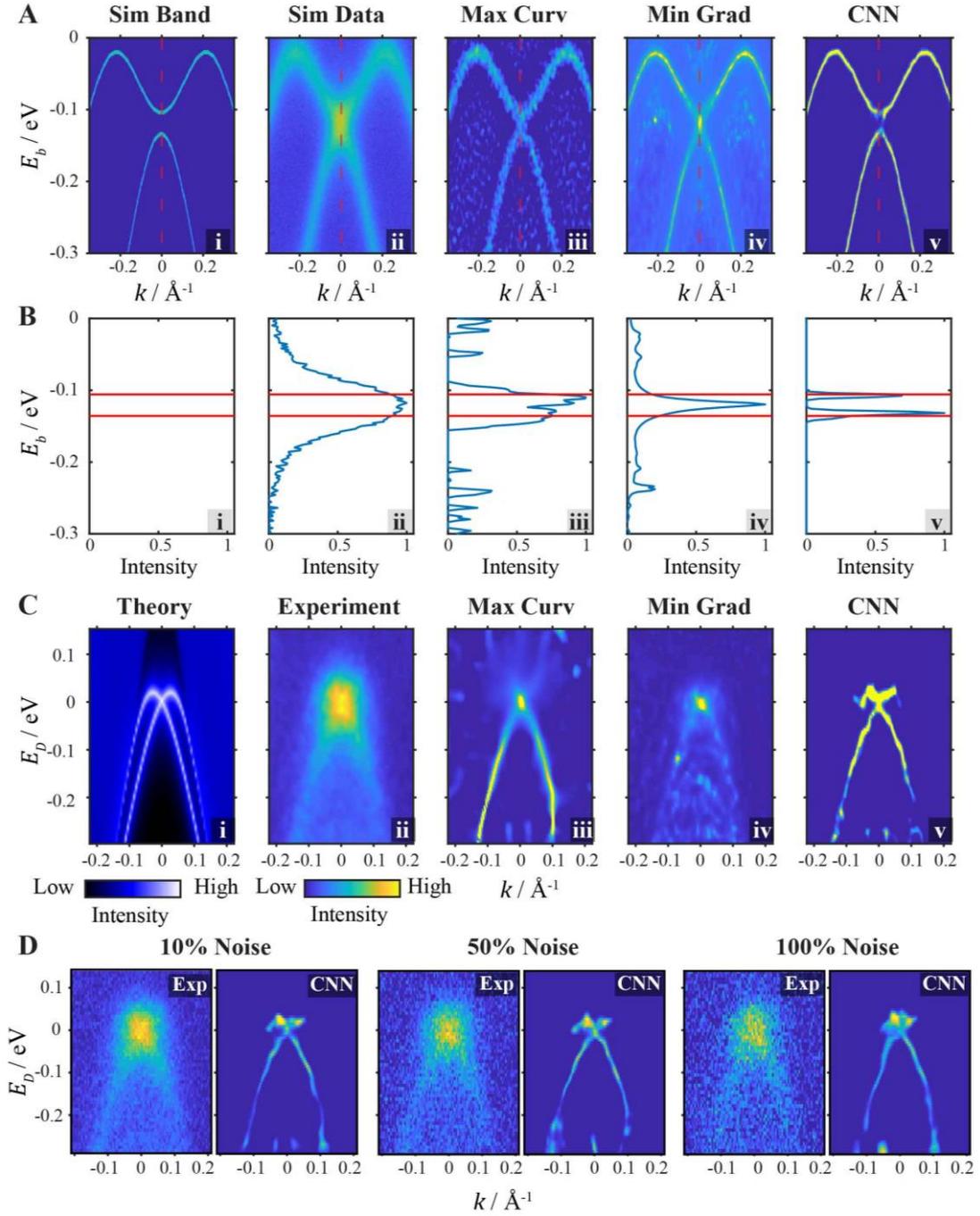

**Fig. 3. Performance analysis of different algorithms. (A)** Performance of different methods on a simulated data. **i.** The simulated energy band dispersion that is used to generate the data. A 30meV gap is at the binding energy Eb~-0.12eV **ii.** The noise-corrupted simulated data used in the following analysis. The small gap is not conceivable directly from the data **iii. – v.** The energy band extraction result based on the simulated data, using the Maximum Curvature method (iii), Minimum Gradient method(iv) and the Convolutional Neural Network method (CNN) (v), respectively. **(B)** The energy distribution curve of k = 0 from the corresponding data in (A) (marked with the red dashed line in (A)). The two red lines (Eb = -0.394 eV and -0.364 eV, respectively) mark the 30meV gap from the original energy band. **(C)** Results for the experiment data of PtSe2 band structure. **i.** Theoretical calculation, showing a crossing at the binding energy $E_D = 0$ eV. **ii.** Experiment data. **iii-v.** The energy band extraction result based on the experiment data, using the method of Maximum Curvature (iii), Minimum Gradient (iv) and CNN (v). **(D)** CNN results with noise-corrupted data. The percentage of noise is the proportion between the variance of the added noise and the experimental data. Negative intensities are padded with zero.

|  | Gap size | Gap size error | Peak width | Contrast |
|---|---|---|---|---|
| Original Data | 30meV | - | 77meV | - |
| CNN | 24meV | 18% | 6meV | 0.99 |
| Maximum Curvature | 25meV | 15% | 23meV | 0.12 |
| Minimum Gradient | N/A | N/A | 16meV | N/A |

**Table 1.** Performance benchmarking of different methods applied to the simulated data

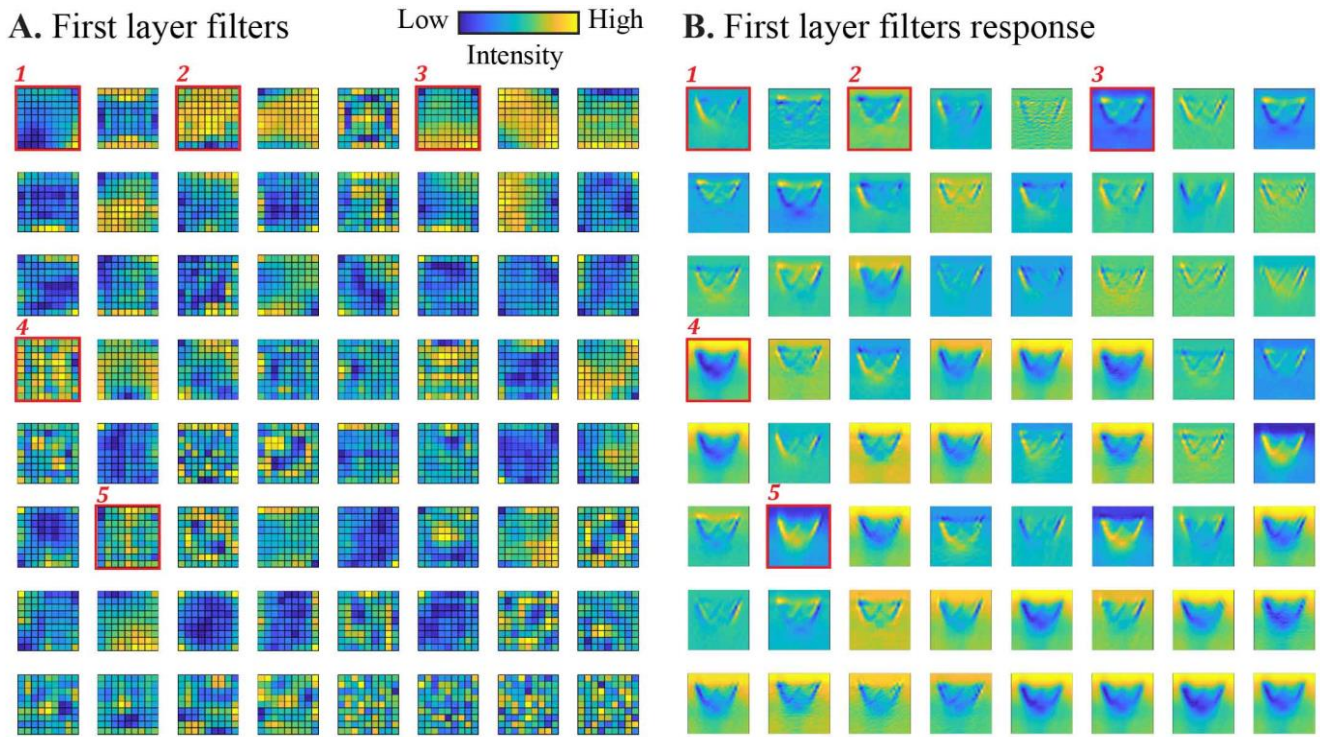

**Fig. 4. The filters and response of the first layer in CNN. (A)** The 64 filters. The filters are arranged in the descending order of variance. **(B)** The response of the filters, which is convolved with the experiment data in Fig. 1.

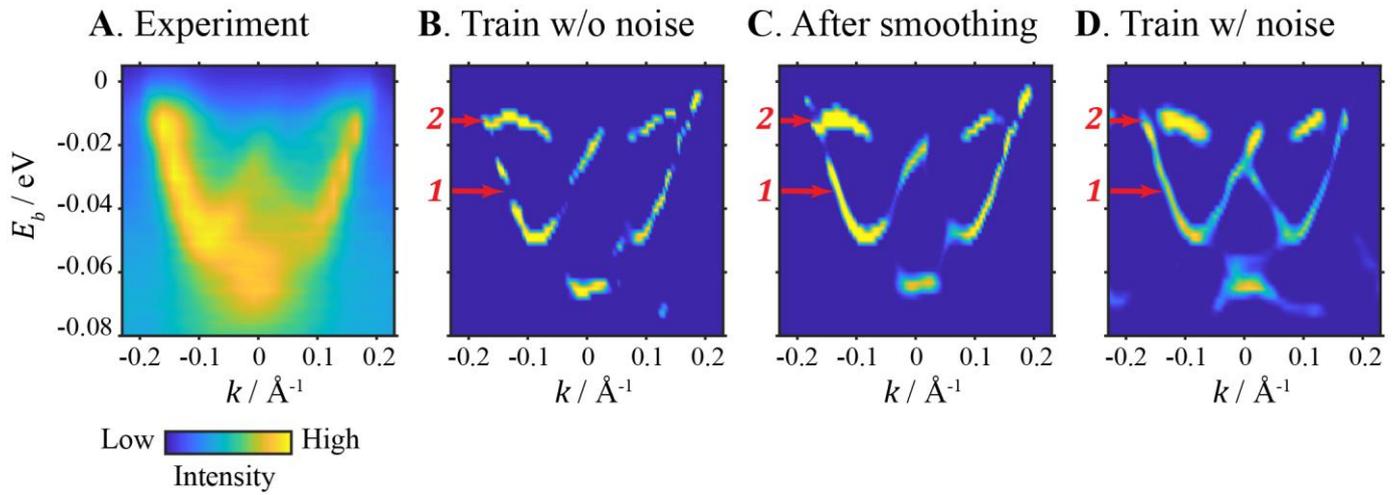

**Fig. 5. The different result of CNN with different training method. (A)** Experiment data in Fig. 1, which is used to produce the subsequent results. **(B)** The energy band extraction result using CNN, which is trained for 30M steps with noise-free simulated data. No pre-processing is used except for interpolation. The extracted band is discontinuous (marked with the red arrow 1). Two bands with small separation are not able to resolve (marked with the red arrow 2). **(C)** The band extraction result using the same training method with (B) but the experiment data is pre-processed with smoothing. The band is more continuous at the position marked by the red arrow 1 but the small feature is still not able to be resolved (marked by the red arrow 2) **(D)** The band extraction result of CNN that is trained for 30M steps with noise-corrupted simulated data. No pre-processing is used except for interpolation. The band is continuous (marked with red arrow 1). The two near bands are clearly separated (marked with red arrow 2).

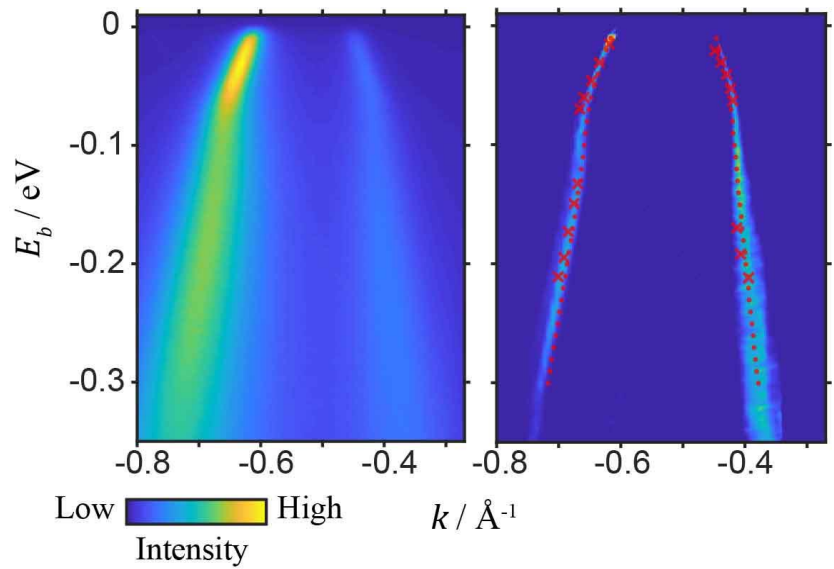

**Fig. 6. CNN result for Bi2212.** (Left) Experiment data. (Right) CNN extracted peak position (false colormap plot). The overlaying red dots are fitted peak positions from MDCs, and the red crosses are manually labelled peak positions from EDCs. To acquire the best band extraction results, the original data is rescaled to two different resolutions to fit-in the tolerant range of the CNN model for the band broadening. The two results are then averaged to produce the final output.